\documentclass[aps,prd,superscriptaddress,showpacs,nofootinbib,eqsecnum,amsfonts,amsmath]{revtex4-1}
\usepackage{epsfig}
\usepackage{longtable}
\usepackage{amssymb}
\usepackage{slashed}

\def\nn{\nonumber}
\def\beq{\begin{equation}}
\def\eeq{\end{equation}}
\def\be{\begin{equation}}
\def\ee{\end{equation}}
\def\bea{\begin{eqnarray}}
\def\eea{\end{eqnarray}}

\def\la{\langle}
\def\ra{\rangle}
\def\ms{\overline{MS}}

\def\t{\tilde }
\def\Lam{\Lambda}

\def\ga{\gamma}
\def\ds{\displaystyle}
\newcommand{\qq}{\langle \bar q q \rangle} 
\newcommand{\lsim}{\raisebox{-0.13cm}{~\shortstack{$<$ \\[-0.07cm] $\sim$}}~}
\newcommand{\gsim}{\raisebox{-0.13cm}{~\shortstack{$>$ \\[-0.07cm] $\sim$}}~}

\begin{document}

\title{The chiral condensate from  renormalization group
optimized perturbation}

\author{Jean-Lo\"{\i}c Kneur}
\author{Andr\'e Neveu}
\affiliation{Laboratoire Charles Coulomb (L2C), UMR 5221 CNRS-Universit\'e de Montpellier, Montpellier, France}

\begin{abstract}
Our recently developed variant of variationnally 
optimized perturbation (OPT), in particular consistently incorporating renormalization group 
properties (RGOPT), is adapted to the calculation of the QCD spectral density of the Dirac operator and the related
chiral quark condensate $\la \bar q q \ra$ in the chiral limit, for $n_f=2$ and $n_f=3$ massless
quarks. The results of successive sequences of approximations at two-, three-, and 
four-loop orders of this modified perturbation, exhibit a remarkable stability. We obtain 
$\la\bar q q\ra^{1/3}_{n_f=2}(2\, {\rm GeV}) = -(0.833-0.845) \bar\Lam_2 $, and 
$ \la\bar q q\ra^{1/3}_{n_f=3}(2\, {\rm GeV}) =  -(0.814-0.838) \bar\Lam_3 $ where the range 
spanned by the first and second numbers (respectively four- and three-loop order results) 
defines our theoretical error, and $\bar\Lam_{n_f}$ is the basic QCD scale in the $\ms$-scheme. 
We obtain a moderate suppression of the chiral condensate when going from $n_f=2$ to $n_f=3$. 
We compare these results with some other recent determinations
from other nonperturbative methods (mainly lattice and spectral sum rules). 
\end{abstract}
\maketitle
\section{Introduction}

The chiral quark condensate $\la \bar q q \ra$ 
plays a central r\^ole in QCD nonperturbative dynamics, being 
together with the pion decay constant the other principal (lowest dimensional) order parameter 
of spontaneous chiral symmetry breaking, $SU(n_f)_L\times SU(n_f)_R \to SU(n_f)_V$ 
for $n_f$ massless quarks, the physically relevant cases being $n_f=2 \, {\rm or}\, 3$. 
It is considered a nonperturbative quantity by excellence, in the sense that it is
trivially vanishing at any finite order of (ordinary) perturbative QCD in the chiral (massless quark) limit, 
as we recall in more details below. 

There is a long history of its determination
from various models and analytic methods, the best known being 
the Gell-Mann--Oakes-Renner (GMOR) relation~\cite{GMOR}, 
relating the quark condensate to the pion mass, the  
decay constant $F_\pi$, and the light current quark masses $m_{u,d}$, typically for the degenerate two-flavor case:
\be
F^2_\pi\,m^2_\pi = -(m_u +m_d) \la \bar u u \ra +{\cal O}(m^2_q)\;.
\label{GMOR}
\ee
Nowadays the light current quark masses can be precisely extracted from lattice simulations~\cite{LattFLAG} or
spectral sum rules (see {\it e.g.} \cite{qqSRlast}),  
giving from using the GMOR relation above an indirect precise determination of the condensate. 
However, as indicated relation (\ref{GMOR}) is valid upon neglecting possible higher order
terms ${\cal O}(m^2_q)$. Indeed, the GMOR relation entails 
explicit chiral symmetry breaking from current quark masses. Since the condensate is more a property of the QCD 
vacuum in the strict chiral limit, it is highly desirable to obtain a possible `first principle'
determination of this dynamical quantity in the strict chiral limit, to disentangle from 
quark current mass effects. 
An early analytic determination was
in the framework of the Nambu--Jona-Lasinio (NJL) model~\cite{NJL} and its various
extensions as a low-energy effective model of QCD, valid at least for the gross features 
of chiral symmetry breaking properties.
In the NJL model, the condensate is evaluated in the strict chiral limit, or taking into account explicit
breaking from small current quark masses, in the leading (large-$N$) approximation~\cite{NJLrev} as a function of the
physical cutoff and other parameters of the model to be fitted from data. 
There are also related other attempts based on analytic methods,
like Schwinger-Dyson equations~\cite{D-S,qqbar_link} typically.
Phenomenological values of the condensate can also be extracted~\cite{qqSR,qqSRlast} 
indirectly from data using spectral QCD sum rule methods~\cite{SVZSSR}, where the quark condensate and other 
higher-dimensional condensates enter as nonperturbative parameters of 
the operator product expansion in inverse powers of momenta. \\

More recently `ab initio' lattice calculations have determined 
the quark condensate by several independent approaches, some actually related to the GMOR relation, or some  
more direct determinations with different methods~\cite{qqlattother} (see \cite{LattFLAG} 
for a review of various recent lattice determinations). In particular 
after early pioneering work~\cite{qqlattSDearly}, there has been more
recently a renewed intense interest in computing on the lattice the spectral density
of the Dirac operator~\cite{qqlattSD,SDlatt_recent}, directly related to the quark condensate through the
Banks-Casher relation~\cite{BanksCasher}. However, while many recent lattice results 
are statistically very precise, lattice determinations rely in the end on extrapolations to the chiral limit, 
often using chiral perturbation theory~\cite{chpt} input for that purpose. 
Earlier general results on the spectral density in the nonperturbative low eigenvalue range  
had been obtained in \cite{SDgen}, and attempts to use chiral 
perturbation theory information was elaborated {\it e.g.} in \cite{SDchpt2}. 
The link between different definitions of the quark condensate,
in particular between the spectral density and the condensate appearing in the operator product expansion (OPE), 
has been carefully discussed in \cite{qqbar_link}. \\

On more phenomenological grounds there are long-standing questions also on the dependence
of the quark condensate on the number of flavors ($n_f=2$ or $n_f=3$ 
for the physically relevant cases). In particular it had been advocated 
that the GMOR relation may receive substantial corrections, and some authors indeed found significant 
suppression of the three-flavor case with respect to the two-flavor case~\cite{qqflav},
which may be attributed to the relatively large explicit chiral breaking from the not-so-small
mass of the strange quark, roughly of order $m_s \lsim \bar\Lam_{QCD}/3$. There are also some hints that chiral
perturbation theory might not converge very well for $n_f=3$~\cite{qqflav,chptn3}, for which value 
lattice simulations also show large discrepancies between different collaborations and methods
(with some results with, and some other results without relative suppression of the $n_f=3$ 
condensate~\cite{LattFLAG}).\\ 

Our recently developed renormalization group optimized perturbation (RGOPT) 
method~\cite{rgopt1,rgopt_Lam,rgopt_alphas} appears particularly adapted to estimate
this quantity, since it gives a nonperturbative sequence of approximations 
starting from a purely perturbative expression. This allows by construction an analytic 
`first principle' calculation of this quantity and gives a nontrivial result 
by construction in the chiral limit,
in contrast with the standard perturbative one (see also \cite{qcd1,qcd2} for earlier attempts in that direction). 
Morever this also gives us a very simple analytic handle
on the exploration of the chiral limit for arbitrary number of flavors ($2$ or $3$ in practice), which
in our framework is simply contained in the known flavor dependence of the first few perturbative coefficients, 
while this appears more difficult at present both for chiral perturbation theory and lattice simulations.\\

The paper is organized as follows. In section II we shortly recall basics of the spectral density
and its Banks-Casher connection with the condensate. In section III we recall the main OPT method and our 
RGOPT version incorporating consistent RG properties. We adapt the RGOPT to the spectral density case 
in section III.C.
Section IV is a digression where we first consider the spectral density RGOPT calculation 
in the Gross-Neveu model, where it can be compared with the exact result for the
fermion condensate in the large-$N$ limit, known from standard methods. Section V deals with the actual
computation of the optimized spectral density in QCD at the three presently available orders 
(two, three and four loops) of the
variationally modified perturbation. Detailed numerical results are presented as well as some comparison
with other recent determinations of the quark condensate. Finally section VI is a conclusion.
\section{Spectral density and the quark condensate}
We shall just recall in this section some rather well-known features 
of the spectral density and its connection with the chiral condensate, 
known as the Banks-Casher relation~\cite{BanksCasher} (see also {\it e.g.}~\cite{SDgen} for
more details), to be exploited below. 
We thus start from the (Euclidean) Dirac operator which formally has eigenvalues $\lambda_n$
and eigenvectors $u_n$,
\be
 {\rm i}\,{\slashed D}\: u_n(x) =  \lambda_n\: u_n(x);\;\;\;{\slashed D} \equiv {\slashed{\partial}} + g\, {\slashed{A}}\, ,
\ee
where ${\slashed D}$ is the covariant derivative operator and $A$ the gluon field. Except for zero modes,
the eigenvectors
come in pairs $\{ u_n(x); \gamma_5\,u_n(x) \}$, 
with respective eigenvalues $\{\lambda_n; -\lambda_n \}$ that depend on $A$.
In the discrete case ({\it i.e.} on a lattice with finite volume $V$), by definition the spectral density is given by
\be
\rho(\lambda) \equiv \frac{1}{V} \la \sum_n \delta(\lambda-\lambda_n^{[A]})\ra\, ,
\label{rho_discrete}
\ee
where $\delta(x)$ here is the Dirac distribution and 
$\la \cdots \ra$ designates averaging over the gauge field configurations,
\be
\la \ra = \int [{\rm d}A] \prod_{i=1}^N \det ({\rm i}{\slashed D}+m)\, .
\ee
The quark condensate is given by
\be
\frac{1}{V} \int_V {\rm d}^4 x \la \bar q(x) q(x)\ra = -2\, \frac{m}{V} \sum_{\lambda_n >0} 
\frac{1}{\lambda^2_n+m^2}.
\ee
Now when $V$ goes to infinity the operator spectrum becomes dense, so that
\be
\la \bar q q  \ra = -2\, m \int_0^\infty {\rm d}\lambda \frac{\rho(\lambda)}{\lambda^2+m^2}\, ,
\label{SD}
\ee
where $\rho(\lambda)$ is the spectral density. \\

The Banks-Casher relation is the $m\to 0$ limit of this, giving
the condensate in the relevant chiral symmetric limit as
\be
\lim_{m\to 0} \la \bar q q  \ra = -\pi \rho(0)\, ,
\label{BC}
\ee
if the spectral density at the origin can be known. This is an intrinsically
nonperturbative quantity, vanishing to all orders of ordinary perturbation, just as
the left-hand side of this last equation. 
Now taking into account that for non-zero fermion masses $m$, 
$\la \bar q q  \ra= \la \bar q q  \ra (m)\equiv -\Sigma(m)$ 
we have from the defining relations 
(\ref{rho_discrete}), (\ref{SD})
the following interesting 
tautology,
\be
\rho(\lambda)=\frac{1}{2\pi} \left[\Sigma({\rm i}\lambda+\epsilon)-
\Sigma({\rm i}\lambda-\epsilon)\right]|_{\epsilon\to 0}\, ,
\label{disc}
\ee
{\it i.e.} $\rho(\lambda)$ is determined by the discontinuities of $\Sigma(m)$ across the imaginary axis.
The interest of this relation is that, when the quark mass is nonzero, 
$\la \bar q q  \ra$ has a purely perturbative series
expansion, known to three-loop order at present, and its discontinuities are simply given
by those coming from the perturbative, purely logarithmic mass dependence. Therefore it makes
sense to calculate a perturbative spectral density using the above relation. Usually
it will be of little use, since taking its $\lambda\to 0$ limit, as relevant
for the true chiral condensate, will only lead to a trivially
vanishing result~\cite{qqbar_link}. But the OPT (and in particular our RGOPT version) 
series modification after performing the variational $\delta$ expansion (see below), 
is precisely the analytic handle giving a nontrivial result for $\lambda \to 0$,
just as it gives a nontrivial result for $m \to 0$ for modified perturbative series with mass dependence.\\

Thus the recipe we shall apply is clear: In a first stage we calculate the purely perturbative
expression of $\rho(\lambda)$ up to  four loops, using the logarithmic discontinuities involved in 
Eq.(\ref{disc}). Then we perform a variational transformation (so-called $\delta$-expansion, defined in next section) 
on the perturbative series, 
and solve appropriate optimization OPT and renormalization group (RG) equations (to be precisely defined in next section) 
to derive a nontrivial, optimized  value of the relation~(\ref{BC}). 
\section{Optimized and RG optimized perturbation (RGOPT)}
\subsection{Standard OPT}
The key feature of the optimized perturbation (OPT) method (appearing in the 
literature under many names and variations~\cite{delta})
is to introduce an extra parameter $0<\delta<1$, interpolating between ${\cal L}_{free}$ and 
${\cal L}_{int}$ for any Lagrangian, in such a way that the mass 
parameter $m$ is traded for an arbitrary trial parameter.
This is perturbatively equivalent
to taking any standard perturbative expansions in the coupling $g(\mu)$, after renormalization
in some given scheme ({\it e.g.} $\ms$-scheme with arbitrary scale $_mu$), reexpanded  
in powers of $\delta$ after substituting,
\be m \to  m\:(1- \delta)^a,\;\; g \to  \delta \, g\,.
\label{subst1}
\ee
Such a procedure is consistent with renormalizability~\cite{gn2,qcd1,qcd2} 
and gauge invariance~\cite{qcd2}, whenever the latter is relevant,
provided of course that the above redefinition of the coupling is performed
consistently for all interaction terms and counterterms
appropriate for renormalizability and gauge invariance, as is the case for QCD.  
In~(\ref{subst1}) we introduced an extra parameter $a$, to reflect a certain freedom 
in the interpolation form, which will be crucial to impose compelling RG
constraints, as discussed below and in our previous work~\cite{rgopt_Lam,rgopt_alphas}.  
Applying~(\ref{subst1}) to a given perturbative expansion 
for a physical quantity $P(m,\lambda)$, reexpanded in $\delta$ 
at order $k$, and taking {\em afterwards} the $\delta\to 1$ limit to recover the original {\em massless} theory,  
leaves a remnant $m$ dependence at any finite $\delta^k$ order. 
The arbitrary mass parameter $m$ is then most conveniently fixed by an optimization (OPT) prescription,
\be
\frac{\partial}{\partial\,m} P^{(k)}(m,g,\delta=1)\vert_{m\equiv \tilde m} = 0\, ,
\label{OPT}
\ee  
which generally determines a nontrivial optimized mass $\t m(g)$, 
having a nonperturbative $g$ dependence, realizing dimensional 
transmutation. (More precisely, for asymptotically free theories, 
the optimized mass is automatically of the order of the basic scale 
$\Lambda \sim \mu\, e^{-1/(b_0\,g)}$, 
in contrast with the original vanishing mass).  
In simpler ($D=1$) models  the procedure may be seen as a particular case of 
``order-dependent mapping''\cite{odm},  
which has been proven\cite{deltaconv} to converge exponentially fast for the $D=1$ $\Phi^4$ oscillator energy levels.  
For higher dimensional $D > 1$ renormalizable models, the behavior at large orders in~$\delta$
is more involved, and no rigorous convergence proof exists, although OPT was shown to partially  
damp the factorially divergent (infrared renormalons) perturbative behavior at large orders~\cite{Bconv}.   
Nevertheless the OPT can give rather successful  approximations for nonperturbative quantities 
beyond mean-field approximations in a large variety of models~\cite{delta,beccrit,bec2}, including studies of 
phase transitions at finite temperatures and densities~\cite{optGN,NJLOPT}.
\subsection{Renormalization group optimized perturbation (RGOPT)} 
In most previous OPT applications~\cite{delta}, the linear $\delta$ expansion is used, namely assuming $a=1$ 
in Eq.~(\ref{subst1}) mainly for simplicity and economy of parameters. However a well-known drawback of this conventional OPT approach is that, 
beyond lowest order, Eq.~(\ref{OPT}) generally gives more and more solutions 
at increasing orders, many being complex-valued, as a result of exactly solving algebraic equations
in $g$ and/or $m$. This problem is typically encountered first 
at two-loop order. In general, without some insight on 
the nonperturbative behavior of the solutions, it can be difficult
to select the right one, and unphysical nonreal solutions at higher orders are embarrassing. 
As it turns out, RG consistency considerations provide a compelling way out, as developed in  
our more recent approach~\cite{rgopt1,rgopt_Lam,rgopt_alphas}, which 
differs crucially from the more conventional OPT based on the linear $\delta$ expansion in two main respects. 
First, it introduces a straightforward combination of OPT and RG  
properties, by  requiring the ($\delta$-modified) expansion to satisfy, in 
addition to the OPT Eq.~(\ref{OPT}), a 
standard RG equation,
\be
\mu\frac{{\rm d}}{{\rm d}\mu} \left(P^{(k)}(m,g,\delta=1)\right) =0\, , 
\label{RG}
\ee 
where the RG operator is defined as usual\footnote{For QCD our normalization is 
$\beta(g)\equiv dg/d\ln\mu = -2b_0 g^2 -2b_1 g^3 +\cdots$,  
$\gamma_m(g) = \gamma_0 g +\gamma_1 g^2 +\cdots$, where $g\equiv 4\pi\alpha_S$. The $b_i$, $\gamma_i$ coefficients
up to four loops are given in~\cite{bgam4loop}.}, 
\be
\mu\frac{{\rm d}}{{\rm d}\mu} =
\mu\frac{\partial}{\partial\mu}+\beta(g)\frac{\partial}{\partial g}-\gamma_m(g)\,m
 \frac{\partial}{\partial m}\, .
 \label{RGop}
\ee 
Note, once combined with Eq.~(\ref{OPT}), the RG equation takes a reduced simple form,
corresponding to a massless theory,
\be
\left[\mu\frac{\partial}{\partial\mu}+\beta(g)\frac{\partial}{\partial g}\right]P^{(k)}(m,g,\delta=1)=0\, .
\label{RGred}
\ee
Thus Eqs.~(\ref{RGred}) and~(\ref{OPT}) together completely fix 
{\em optimized} $m\equiv \t m$ and $g\equiv \t g$ values. \\

Remark indeed that 
RG invariance is in general spoiled after the rather 
drastic modification from (\ref{subst1}), reshuffling interaction and free terms from the original perturbative expansion. 
This feature has seldom been considered and appreciated in previous applications of 
the OPT based on the linear $\delta$-expansion method to renormalizable theories. 
Thus RG invariance has to be restored in some manner, accordingly Eq.~(\ref{RGred}) gives an additional non-trivial constraint.
Intuitively, just as the stationary point OPT solutions from Eq.~(\ref{OPT}) are expected to give sensible approximations, 
at successive orders, to the actually massless theory, one similarly expects that combining the OPT 
with the RG solutions of~(\ref{RGred}) should further give a sensible sequence of 
best approximations to the exactly scale
invariant all order result.
(N.B.: An earlier way of reconciling the $\delta$ expansion with RG properties
was used in~\cite{gn2,qcd1,qcd2}: Schematically it  
amounted to resumming the $\delta$-expansion to all orders, which can be done in practice only for 
the pure RG dependence up to two-loop order. 
These resummations came as rather complicated 
integral representations, rendering difficult  
generalizations to higher orders, other physical quantities or other models of interest. 
In contrast the purely perturbative procedure of imposing Eqs.~(\ref{OPT}), (\ref{RGred})
is a considerable shortcut, straightforward to apply to any 
model, being based solely on purely perturbative expansions.)

Yet applying Eq.~(\ref{OPT}), (\ref{RGred}) without further insight still gives multiple solutions at increasing orders.  
So we proposed~\cite{rgopt_Lam,rgopt_alphas} a compelling selection criterion, by retaining only the 
branch solution(s) $g(m)$ (or equivalently $m(g)$) 
{\em continuously matching} the standard perturbative (asymptotically free (AF) RG behavior in the QCD case) 
for vanishing coupling, namely, 
\be
\t g (\mu \gg \t m) \sim (2b_0 \ln \frac{\mu}{\t m})^{-1} +{\cal O}( (\ln \frac{\mu}{\t m})^{-2})\, .
\label{rgasympt}
\ee
Now the crucial observation is that requiring at least one of the solutions of Eq.~(\ref{RGred}) to satisfy (\ref{rgasympt})  
implies a strong necessary condition on the 
basic interpolation (\ref{subst1}), fixing the exponent $a$ uniquely in terms of the universal (scheme-independent) 
first order RG coefficients~\cite{rgopt_Lam,rgopt_alphas}:
\be
a\equiv \frac{\ga_0}{2b_0}\, ,
\label{acrit}
\ee
which is the second important difference of the present RGOPT with respect to the standard OPT~\footnote{The important role of 
the anomalous dimension $\ga_0/(2b_0)$ appeared also   
in our earlier constructions resumming the RG dependence of the $\delta$-expansion~\cite{gn2,qcd1,qcd2,Bconv}, although it had not been recognized at that time 
as a crucial RG consistency property of lowest $\delta$ orders.}.
For the critical value (\ref{acrit}), Eq.~(\ref{RGred}) is in fact exactly satisfied at the lowest $\delta^0$ order, 
therefore giving no further
constraint. At higher $\delta$ orders, (\ref{acrit}) implies that one at least of both the RG and OPT solutions fulfills Eq.~(\ref{rgasympt}), and solutions 
with this behavior are essentially unique (although not necessarily) at a given perturbative order. 
Moreover, taking (\ref{acrit}) drastically improves the convergence of the method: More precisely 
the known nonperturbative result of generic pure RG-resummed expressions are obtained exactly from RGOPT at the very first $\delta$-order, while
the convergence of the conventional OPT with $a=1$ is not clear or very slow, if any
(see Sec.~III.C of ref.~\cite{rgopt_alphas} for details).

The criterion (\ref{rgasympt}) can easily be generalized to any model, even nonasymptotically free ones, 
by similarly selecting those optimized solutions that simply match 
the standard perturbative behavior for small coupling values. Thus clearly 
the resulting unique critical value like in (\ref{acrit}) is valid
for any model with its appropriate RG coefficients. 
For the QCD spectral density, as we will see below the equivalent of the criteria (\ref{rgasympt}) indeed selects a unique 
solution at a given order for both the RG and OPT equations, at least up to the four-loop order 
(as was also the case for the pion decay constant~\cite{rgopt_alphas}). 

Incidentally, a connection of the exponent $a$ with RG anomalous dimensions/critical exponents 
had also been established previously in a different context, 
in the  $D=3$ $\Phi^4$ model for the Bose-Einstein condensate (BEC) 
critical temperature shift, by two independent approaches~\cite{beccrit,bec2}, where for this model
it also leads to real OPT solutions~\cite{bec2}. Indeed in ref.~\cite{beccrit,HKcrit} it was convincingly argued, based on
critical behavior considerations, that 
the OPT can only converge if an appropriate Wegner critical exponent is used in the interpolation (\ref{subst1}), which appears
quite similar to our criterion (\ref{acrit}). Note however that (\ref{acrit}) is identified  
exactly from the known first-order RG coefficients, thus valid for 
any model, while in \cite{HKcrit} the analoguous exponent was determined more approximately  
by looking for a plateau in the variational parameter dependence. In any case, from these examples,
it is established that it is necessary for the OPT method to give useful results
to have in general an exponent $a$ in (\ref{subst1}) differing from 1 in a well-defined way. 

Coming back to the present OPT and RG Eqs.~(\ref{OPT}), (\ref{RGred}), beyond lowest orders 
AF-compatible solutions with behavior (\ref{rgasympt}) are however not necessarily real in general. 
A rather simple way out is to further exploit the RG freedom, considering 
a perturbative renormalization scheme change to attempt to 
recover  RGOPT solutions both AF-compatible and real~\cite{rgopt_alphas}. 
In the present case of the spectral density, this extra complication is not even necessary, at least
up to the highest order here studied (four loops): We shall find also that the unique AF-compatible RG and OPT solution remains real. 

\subsection{RGOPT for the spectral density}
Formally a generic pertubative expansion for the condensate reads typically
\be
\la \bar q q \ra_{\mbox{pert}} =  m^3 \sum_{p} g^p \sum_{k=0}^p f_{pk} \ln^{p-k} (\frac{m}{\mu})\, ,
\label{genpert}
\ee
where the $f_{pk}$ coefficients are determined by RG properties from lowest orders for $k < p$.
According to Eq.~(\ref{disc}), calculating the (perturbative) spectral density 
formally involves calculating all logarithmic discontinuities. 
This is conveniently given  by taking in any typical perturbative expansion 
like~(\ref{genpert}), all nonlogarithmic terms to zero, those trivially having no discontinuities,
while replacing all powers of logarithms, using $m\to {\rm i}|\lambda|$ etc.,  as 
\be
\ln^n \left(\frac{m}{\mu}\right) = \frac{1}{2^n} \ln^n \left(\frac{m^2}{\mu^2}\right)\to  \frac{1}{2^n} \frac{1}{2{\rm i} \pi}
\left[ \left(2\ln \frac{|\lambda|}{\mu}+{\rm i}\pi\right)^n -\left(2\ln \frac{|\lambda|}{\mu}-{\rm i}\pi\right)^n \right]\, ,
\label{discgen}
\ee
leading to the following simple substitution rules for the first few terms
\be
\ln \left(\frac{m}{\mu}\right) \to 1/2;\;\; 
\ln^2 \left(\frac{m}{\mu}\right) \to \ln \frac{|\lambda|}{\mu};\;\; 
\ln^3 \left(\frac{m}{\mu}\right) \to \frac{3}{2} \ln^2 \frac{|\lambda|}{\mu} -\frac{\pi^2}{8}\, ,
\label{disc3}
\ee
and so on 
(note the appearance of nonlogarithmic $\sim \pi^2$ terms starting at order $\ln^3 m$).
This gives a perturbative expression of the spectral density of the 
generic form
\be
\rho_{pert}(|\lambda|,g) =  |\lambda|^3 \sum_{p\ge 1} g^p \sum_{k=0}^p f^{SD}_{pk} \ln^{p-k} 
(\frac{|\lambda|}{\mu})\, ,
\label{SDgen}
\ee
where the determination of the coefficients $f^{SD}_{pk}$ follows 
from the above relations (\ref{discgen}).\\
To obtain the RG equation for $\rho(g,\lambda)$, we
use the defining integral representation of the spectral density in (\ref{SD})
and the basic algebraic identity
\be
\frac{\partial}{\partial m} \frac{m}{\lambda^2 + m^2} = 
-\frac{\partial}{\partial \lambda} \frac{\lambda}{\lambda^2 + m^2}\, .
\label{m->lambda}
\ee
Throwing away surface terms
in partial integrations as usual in the spirit of dimensional regularization, one thus finds
that $\rho(\lambda)$ actually obeys the same RG equation as $\la \bar q q \ra$,
with  $\partial m$ replaced by $\partial \lambda$,
\be
\left[\mu\frac{\partial}{\partial\mu}+\beta(g)\frac{\partial}{\partial g}-\gamma_m(g)\,\lambda
 \frac{\partial}{\partial\lambda} -\gamma_m(g) \right]\rho(g,\lambda)=0.
\label{RGlam}
\ee 
One can next proceed to the modification of the resulting perturbative series $\rho(\lambda,g)$ 
as implied by the $\delta$-expansion, now, from Eq.~(\ref{m->lambda}) clearly applied 
not on the original mass but on the
spectral value\footnote{We adopt in the following
the notation $\lambda\equiv|\lambda|$ since it is necessarily positive.} $\lambda$:
\be
\lambda\to \lambda (1-\delta)^a\, \;\;g\to \delta\,g\, .
\label{substlam}
\ee
Optimizing perturbation theory means that the derivative with respect to $m$ of
\be
\sum_{n=0}^\infty \frac{(-1)^n}{n!}m^n\left(\frac{\partial}{\partial m}\right)^n \la\bar q q \ra\, ,
\label{OPT2}
\ee
being formally zero\footnote{For simplicity, we have set $a$ to one in this equation.}, 
one should obtain a good approximation for the value at $m=0$ of $\la\bar q q\ra $ at finite order by 
setting to zero the derivative of a finite number of terms of this series, see Eq.~(\ref{OPT}). Using Eq.~(\ref{m->lambda}), this mass
optimization on $\la\bar q q \ra$ thus translates into an optimization
of the spectral density with respect to $\lambda$,
\be
\frac{\partial \rho^{(k)}(\lambda,g)}{\partial \lambda}= 0\, ,
\label{OPTlam}
\ee
at successive $\delta^k$ order.
%
\section{Lessons from the Gross-Neveu model}
The fermion condensate can also be defined from the spectral density
for the $D=1+1$ $O(N)$ Gross-Neveu (GN) model~\cite{GN}. This will give us a very useful guidance for the more elaborate
QCD case below, and we also set up some formulas actually generically valid for both the GN model
and QCD.\\

We start from the known expression of the vacuum
energy evaluated in the large-$N$ limit~\cite{gn2}
after all necessary mass, coupling, and vacuum energy (additive) renormalizations, 
in terms of the explicit mass $m$ and mass gap $M(m,g)$,
\be
E_{GN} = -\left(\frac{N}{4\pi}\right)\,\left( M^2(m,g) +2\frac{m}{g} M(m,g)\right)\, ,
\label{vacGN}
\ee
where the mass gap is defined in compact form as 
\be
M(m,g) = m \left(1+g \ln \frac{M}{\mu} \right)^{-1}\, ,
\label{MLNpert}
\ee
where $m\equiv m(\mu)$ and $g\equiv g(\mu)$ are the renormalized mass and coupling
in the $\ms$ scheme (after convenienly rescaling the original coupling defined by 
$ (1/2) g^2_{GN} (\bar \Psi \Psi)^2 $ as $g^2_{GN} N/\pi \equiv g$). 
The fermion condensate is formally given by the derivative with respect to $m$ 
of the vacuum energy, giving simply after some algebra
\be
\la \bar \psi \psi \ra_{GN}(m,g) \equiv -\left(\frac{N}{2\pi}\right)\,\frac{M(m,g)}{g}\, .
\label{qqGN}
\ee
This means that up to a trivial overall factor, the fermion condensate is directly related to the mass gap,
as intuitively expected also from mean-field arguments. Eq.~(\ref{qqGN}) has a well-defined nontrivial
perturbative expansion to arbitrary order,
\bea 
\ds -\left(\frac{2\pi}{N}\right) \la \bar \psi \psi \ra_{GN}(m,g) \equiv  -\la \bar \Psi \Psi \ra_{GN}
\simeq m \left(\frac{1}{g} - L_m +g L_m (L_m+1)
  -\frac{1}{2} L_m (L_m+2)(2L_m+1) g^2 +{\cal O}(g^3) \right) ,
  \label{qqGNpert}
\eea
where $L_m\equiv \ln m/\mu$ (and for convenience we redefined 
the condensate by a trivial rescaling). From properties of the implicit $M(m)$ defined by~(\ref{MLNpert}) and its reciprocal function $m(M)$,  
one can establish~\cite{gn2} that $M(m)\to \Lam \equiv \mu e^{-1/g}$ for $m\to 0$, 
which translates here into the simple relation
\be
-\la \bar \Psi \Psi \ra_{GN}(m\to 0) = \frac{\Lam}{g}\, ,
\label{qqGNexact}
\ee
which provides a consistent bridge between the massive
and massless case. But deriving this needs the knowledge of the 
all-order expression~(\ref{MLNpert}), only known exactly in the large-$N$ limit.\\

Now alternatively, performing the substitution\footnote{Taking $a=1$ as appropriate for the large-$N$ GN model.}~(\ref{subst1}), expanding at order $\delta^p$,
setting $\delta$ to one, and optimizing the resulting expression with Eqs.~(\ref{OPT})
and (\ref{RGred}) gives the exact result~(\ref{qqGNexact})
at any order in $\delta^p$, at optimized coupling and mass values,
\be
\t g =1 ;\;\;\t L_m =-1\, ,
\label{gLGN}
\ee
just as in the mass gap case~\cite{rgopt1}. This is not very surprising, in view
of the rather trivial  relation~(\ref{qqGNexact}) between the mass gap and the condensate 
in the large-$N$ limit.\\

However here we shall try to obtain this large-$N$ result in an indirect way, using the spectral density, the aim being evidently to test on the exactly known result~(\ref{qqGNexact}) the possibility of calculating a spectral density and of estimating
its reliability from the first few perturbative orders,
in order to subsequently apply the same
procedure in the more challenging QCD case.\\

Thus from~(\ref{qqGNpert}) and using $m\to {\rm i}|\lambda|$ and relations~(\ref{discgen}) 
the perturbative expression of the spectral density, 
for instance restricted up to four-loop order $g^3$, follows as
\bea
\rho_{GN}^{pert}(\lambda,g) \simeq  \lambda &&\left(-\frac{1}{2} +g \left(L_\lambda +\frac{1}{2}\right)
  -\frac{g^2}{8} \left(12 L^2_\lambda +20L_\lambda +4-\pi^2\right) + \right. \nn \\
&& \left.  \frac{g^3}{24} \left(48 L^3_\lambda +156 L^2_\lambda +(108-12\pi^2) L_\lambda + 12 -13\pi^2\right)
+  {\cal O}(g^4) \right)\, ,
  \label{rhoGNpert}
\eea
where now $L_\lambda\equiv \ln \lambda/\mu$.
We can then proceed by applying~(\ref{substlam}), expanded to order $\delta^k$, then taking $\delta\to 1$, 
and finally applying on the resulting expression the OPT~(\ref{OPTlam}) and RG~(\ref{RGlam}) Eqs. 
In practice we shall proceed to relatively high perturbative orders, 
since the exact expression~(\ref{MLNpert}) may be formally expanded to arbitrary order. \\ 
Thus with those obvious replacements
we can proceed first with the $\delta$-expansion~(\ref{substlam}) operating on $\lambda$ and $g$,
and taking the relevant value of the exponent for the large-$N$ case, 
$a\equiv \gamma_0/(2b_0) =1 $ in~(\ref{subst1}).
At first nontrivial $\delta^1$ order we simply obtain
\be
\rho^{\delta^1}(\delta\to 1; \lambda, g)= \lambda g \left(\frac{1}{2}+L_\lambda\right)\, ,
\label{rhoGNd1}
\ee
from which the OPT~(\ref{OPTlam}) and RG~(\ref{RGlam})
give the unique solution,
\be
\tilde L_\lambda \left(\equiv \ln \frac{|\lambda|}{\mu}\right) = -\frac{3}{2}; \;\; \tilde g= \frac{1}{2}\, ,
\ee
which plugged back into~(\ref{rhoGNd1}) gives the final result, using also the 
Banks-Casher relation~(\ref{BC}),
\be
\t g\,\la \bar \Psi \Psi \ra_{GN}(m\to 0)/\Lam \equiv -\pi \t g \rho(\tilde \lambda,\tilde g)/\Lam 
=  -\frac{\pi}{4}\:\sqrt e \simeq -1.2949\, ,
\ee
which is to be compared to the exact result $g\, \la \bar \Psi \Psi \ra_{GN}(m\to 0)/\Lam = -1$
in this normalization. We then proceed to rather high perturbative order, which in practice  
becomes relatively involved algebraically but can be easily handled with computing softwares 
like ${{Mathematica}}^\circledR$\cite{mathematica}).\\
\begin{table}[h!]
\begin{center}
\caption[long]{RGOPT $\frac{-g \la \bar \Psi \Psi \ra_{GN}}{\Lam}$ and corresponding optimized coupling 
$\t g$ and (logarithm of) optimized eigenvalue $\ln \t \lambda/\mu$ 
results at successive  orders in $\delta$.} 
\begin{tabular}{|l||c|c|c|c|}
\hline
$\delta^k$ order  &  $\frac{- g \la \bar \Psi \Psi \ra_{GN}}{\Lam}$ & $\t g$ 
& $ \ln \frac{\t \lambda}{\mu}$  \\
\hline
1 &   $1.295 $    & $\frac{1}{2}$ & $-\frac{3}{2} $ \\
2  & $ 0.984 $ & $0.398 $  & $-2.547$  \\
3   & $0.897$ & $0.335$ & $ -3.278$\\
4   & $1.081$ & $0.399$ & $-1.373 $\\
5   & $0.924$ & $0.394$ & $-1.919 $\\
6   & $0.877$ & $0.372$ & $-2.337 $\\
7   & $0.980$ & $0.386$ & $ -1.332$\\
8   & $0.903$ & $0.389$ & $-1.716 $\\
9   & $1.065$ & $0.358$ & $-0.928 $\\
10  & $0.934$ & $0.383$ & $-1.312 $ \\
11  & $0.894$ & $0.385$ & $-1.613 $ \\
12  & $ 0.978 $ & $ 0.364$  & $-1.001$  \\
13   & $0.972$ & $0.349$ & $-3.77 $\\
14   & $1.013$ & $0.339$ & $-4.049 $\\
15   & $0.989$ & $0.403$ & $-2.902 $\\
16   & $1.027$ & $0.391$ & $-3.107 $\\
17   & $0.978$ & $0.428$ & $-2.405 $\\
18   & $1.013$ & $0.420$ & $-2.574 $\\
\hline
\end{tabular}
\label{GNrho20}
\end{center}
\end{table}
The results up to order $g^{18}$ (beyond which numerics become really tedious) are given in Table~\ref{GNrho20} 
(where we give for convenience the scale invariant condensate $\t g \la \bar \Psi \Psi \ra$ values). Actually, starting at order
$g^3$, there are more than one real solution. We show here those
solutions which are unambiguously determined to be the closest to the standard 
AF perturbative behavior, $L_\lambda \simeq -1/g +{\cal O}(1)$ for $g \to 0$,  by analogy with
the exact value  $-\t g \t L_m\equiv 1$ obtained for the simpler mass gap case. One can see a regular pattern for
the values of the optimized coupling for such solutions, with $\t g \simeq 0.4$, while the optimal spectral 
parameter $\t L_\lambda$ is less stable with wider variations at successive orders. 

These approximate results at successive orders when optimizing the spectral transform are clearly in contrast with 
those obtained from optimizing directly the original perturbative expansion,
where as explained above the AF compatible solution always gives the exact result for the condensate, with the 
corresponding optimal coupling and mass values~(\ref{gLGN}) obtained 
already at first and all successive orders.
Here we see in Table~\ref{GNrho20} that the second order is very close, less than 2\%, from the exact result, but
at higher orders the solutions exhibit a rather slow  
empirical convergence, with an oscillating behavior towards the exact large-$N$ limit result. \\
This slow convergence can be essentially  
traced to the effect of numerous factors $\pi^{2p}$, $p=1,...n/2$ 
appearing at perturbative orders
$g^n$, $g^{n+1}$ for $n\ge2$ from the discontinuities. These terms clearly spoil 
the originally neat simple form of the large-$N$ resummed mass in~(\ref{MLNpert}). Moreover, 
starting at order $g^3$  these $\pi^{2p}$ terms come with larger 
coefficients {\em relative} to the other non-logarithmic coefficients, originating from the 
$\ln m$ coefficients in the original perturbation~(\ref{MLNpert}),~(\ref{rhoGNpert}) which 
are roughly all of the same order ${\cal O}(1)$. More precisely 
inspecting~(\ref{rhoGNpert}) one can see that at order 
$g^2$ the $-\pi^2/8$ contribution is numerically very roughly twice as large
as the other non-logarithmic coefficient ($1/2$), while at order
$g^3$ the relevant terms to be compared are the two last ones: $12/24$ and $-(13/24)\pi^2$, so the latter is an order
of magnitude larger than the former. This explains the very good result at order $g^2$ and also 
the degraded result at the next $g^3$ order. A similar behavior is observed at higher orders, until 
at sufficiently high order the original perturbative coefficients of $\ln m$ also start to grow fastly, such that
a balance with the $\pi^{2p}$ contributions can again occur, with the (slow) convergence observed. 
Indeed the spectral parameter optimization~(\ref{OPTlam}) tends to damp these relative $\pi^{2p}$
contributions: For instance the combined contributions of all $L_\lambda$-dependent terms 
for the optimal value $\t L_\lambda\sim -3.28$ at order $g^3$ in Table~\ref{GNrho20} almost cancel the 
large $\pi^2$ term. But since the latter terms are growing with the order, it is not surprising that
the optimal $\t L_\lambda$ values are not much stable in Table~\ref{GNrho20}.

Interestingly however, if instead of taking the exact results at a given order,
we consider a well-defined approximation, by keeping, 
at growing $\delta^n$ order, {\em only} $\pi^{2p}$ terms with a fixed maximal power, 
then there is a higher $\delta^n$ order at which 
one recovers the simple exact RGOPT solution\footnote{This nice property may appear somewhat artificial but it
can be explained more rigorously: As observed in~\cite{rgopt1} 
the mass gap $M(m,g)$ at a given (sufficiently high) order $\delta^n$, 
has flat optima roughly at order $\sim n/2$ ({\it i.e.} its $n/2$th derivative  with respect 
to $m$ vanishes). Now for the spectral density, $\pi^{2p}$ terms 
appearing at order $g^p$ arise from~(\ref{disc})
as the coefficient of the (logarithmic) derivatives $\partial/\partial \ln m$ of order $p$: 
So {\em if} discarding terms of higher power $\pi^{2q}, q >p$, 
there is necessarily a fixed higher order at which the cancellation
of all $\pi^{2p}$ terms occur.}: 
$\t g=1, \t L_\lambda =-1, \la \bar \Psi \Psi \ra_{GN}=-\Lam$. 
For instance, keeping only $\pi^2$  and $\pi^4$ terms (the latter 
appearing first at order $g^4$) and increasing the $\delta^k$ order, 
the exact solution is recovered at order $\delta^8$. 
This cancellation mechanism  thus indicates 
that the ``maximal convergence" properties of RGOPT, specific of the simpler 
GN model mass gap expression in~(\ref{MLNpert})~\cite{rgopt1}, 
are not completely lost within the perturbative spectral density, but rather hidden, being 
obstructed by the more involved perturbative 
coefficients.   
Those remarks are to be kept in mind when comparing with the QCD case below.
\section{Determination of the QCD quark condensate}
\subsection{Perturbative three-loop quark condensate}
\begin{figure}[h!]
\epsfig{figure=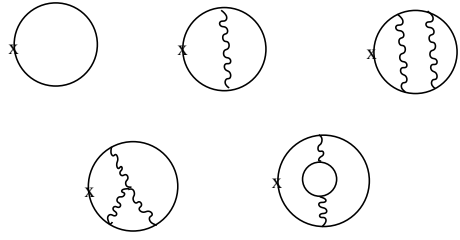,width=6cm}
\caption{Samples of standard perturbative QCD contributions to the chiral condensate 
up to three-loop order. The cross denotes a mass insertion.}
\label{qqpert}
\end{figure}
The perturbative expansion of the QCD quark condensate for a nonzero quark mass can be calculated
systematically from the related vacuum energy graphs. 
A few representative Feynman graphs contributions 
at successive orders up to three-loop order are illustrated in Fig.~\ref{qqpert} (there
are evidently a few more three-loop contributions not shown here). 
Note that the one-loop 
order is ${\cal O}(1)={\cal O}(g^0)$. The two-loop contributions were computed 
long ago~\cite{vac_anom2} 
and the three-loop ones in \cite{vac_anom3}.
The 
three-loop order result in the $\ms$-scheme reads explicitly
\be
m\,\la \bar q q \ra_{QCD}(m,g) =
 \frac{3}{2\pi^2}\,m^4 \: 
\left(\frac{1}{2}-L_m +\frac{g}{\pi^2} (L_m^2 -\frac{5}{6} L_m +\frac{5}{12}) +(\frac{g}{16\pi^2})^2\; 
q_3(\lambda, n_f)\right)\, ,
\label{qqQCDpert}
\ee
where $m\equiv m(\mu)$ and $g\equiv 4\pi\alpha_S(\mu)$ are the running mass and coupling in the $\ms$ scheme, 
and the three-loop coefficient 
reads\footnote{The originally calculated expressions in refs.~\cite{vac_anom3} are
given for arbitrary $N_c$ colors, $n_h$  massive quarks and $n_l$ {\em massless} quarks entering at 
three-loop order. In our context $m$ is the ($n_f$-degenerate) light quark mass 
and its precise mass dependence is what is relevant for the optimization procedure, so one 
should trace properly the full $n_l$, $n_h$ dependence,  
and take $n_l=0,\, n_h\equiv n_f$ with $n_f=2\, (3)$ for the $SU(2)$ (resp. $SU(3)$) 
case.}~\cite{vac_anom3}
\bea
 q_3(\lambda, n_f)=&&\frac{1}{27}\left(6185 - 768\, a_4 - 32\ln^4 2 + 192 \ln^2 2 z_2 + 504 \,z_3 
 +  (672\, z_3 -750)\, n_f + 528\, z_4 \right) \nn \\ 
&&  +\, ( 52\, n_f -\frac{4406}{9} + \frac{32}{3}\, z_3) L_m - 
   \frac{32}{9}\,( 5\, n_f -141 ) L^2_m + 
 \frac{32}{9}\,(2\, n_f -81) L^3_m \, ,
\eea
where $z_i\equiv \zeta(i)$ and $a_4 = {\rm Li}_4(1/2)$ .

The calculation in dimensional regularization
of~(\ref{qqQCDpert}) actually still contains divergent terms needing 
extra subtraction after mass and coupling renormalizations in the
$\ms$ scheme.   
The correct procedure to obtain a RG invariant finite expression when subtracting those 
divergences consistently is well known in the
standard renormalization of composite operators with mixing~\cite{Collins}.
We can define~\cite{qcd2} the needed subtraction as a perturbative series,
\be
{\rm sub}(g,m) \equiv \frac{m^4}{g} \sum_{i\ge 0} s_i  g^{i}\, ,
\label{sub}
\ee
with coefficients determined order by order by
\be
\mu \frac{\rm d}{{\rm d}\mu}{\rm sub}(g,m) \equiv {\rm Remnant}(g,m)=\mu \frac{\rm d}{{\rm d}\mu}[m \la \bar q q \ra({\rm pert})|_{\rm finite}]\, ,
\label{rganom}
\ee 
where the remnant part is obtained by applying the RG operator Eq.~(\ref{RGop}) to the 
finite expression~(\ref{qqQCDpert}), not separately RG invariant. Eq.~(\ref{sub}) does 
not contain any $\ln m/\mu$ terms and necessarily 
starts with a $s_0/g$ term to be consistent with RG invariance properties.  
To obtain RG invariance at order $g^k$, fixing $s_k$ in (\ref{sub}), one needs 
knowledge of the coefficient of the $\ln m$ term 
(equivalently the coefficient of $1/\epsilon$ in dimensional regularization) at order  $g^{k+1}$.
Concerning the condensate 
the extra contribution to the RG equation~(\ref{rganom}) is the 
so-called anomalous dimension of the QCD (quark) vacuum energy, entering the renormalization
procedure of the $m \qq$ operator due to mixing with $m^4\times 1$, and also given explicitly to three-loop order
in \cite{vac_anom3}.  
The $s_i$ coefficients can be expressed in terms of RG coefficients and other 
terms using RG properties. In compact form for completeness in our normalization they read\footnote{The expression of $s_3$
here approximated to $10^{-3}$ relative uncertainty, largely sufficient for our purpose, needs the knowledge
of the four-loop $\ln m/\mu$ coefficient, or alternatively the four-loop vacuum energy anomalous dimension. 
The latter, not explicitly available in the published literature so far, has been kindly provided to us 
by K. Chetyrkin and P. Maier~\cite{vac_anom4} from a related work.} ,
\bea
&& s_0 = \frac{1}{4\pi^2(b_0-2\, \gamma_0)}\, , \nn \\
&& s_1= -\frac{1}{12\pi^2} +\frac{1}{4} \frac{b_1-2\, \gamma_1}{b_0-2\,\gamma_0}\, , \nn \\
&& s_2 =  -\frac{ (112077 + 24519\, n_f + 2101\, n_f^2 + 576(15 - 58\, n_f)\, z_3)}
{1152\, \pi^4 (-81 + 2\, n_f)(15 + 2\, n_f)}\, , \nn \\
&& s_3 \simeq  \frac{ 4.7\, 10^{-10} a_4 (-81 + 2 n_f)(-4.615 + n_f)(15 + 2 n_f) 
+ 2.47 +1.275 n_f +0.303 n^2_f -0.019 n^3_f +3.1\,10^{-4}n^4_f }
{(-57 + 2\, n_f)(-81 + 2\, n_f)(15 + 2\, n_f)}\, . \nn \\
\label{s_i}
\eea
\subsection{Perturbative spectral density}
One crucial advantage of using the spectral density with the Banks-Casher 
relation~(\ref{BC}) is that it gives a direct access to the QCD condensate in the chiral limit, unlike 
the original direct RG invariant expression $m \la \bar q q\ra$ in (\ref{qqQCDpert}), where 
the condensate is being screened by the mass for $m \to 0$. Indeed, 
note that a direct RGOPT optimisation in the GN model of the corresponding expression 
$m \la \bar\Psi \Psi \ra $ gives an exactly vanishing result consistently at any orders~\cite{rgopt_alphas}
(even though the optimized GN mass is clearly nonvanishing, $\t m =\Lam$).

Taking thus the logarithmic discontinuities according to Eqs.~(\ref{disc}),(\ref{discgen}) gives us
the perturbative spectral density up to three-loop order,
\be
-\rho^{\ms}_{QCD}(\lambda)=  \frac{3}{2\pi^2}\,\lambda^3\:  
\left(-\frac{1}{2} +\frac{g}{\pi^2} \left(L_\lambda -\frac{5}{12}\right) +(\frac{g}{16\pi^2})^2 \;
q^{SD}_3(\lambda,n_f)\right)\, ,
\label{SDqq3l}
\ee
where now $L_\lambda\equiv \ln (|\lambda|/\mu)$ and 
\bea
q^{SD}_3(\lambda,n_f) = && \frac{1}{2}\,(52\, n_f -\frac{4406}{9} + \frac{32}{3}\, z_3)  
- \frac{32}{9} ( 5\, n_f -141) L_\lambda \nn \\
&& + \frac{32}{9} (2\, n_f -81 )(\frac{3}{2} L^2_\lambda -\frac{\pi^2}{8}) \, .
\label{q3lSD}
\eea
Note that the $\pi^2$  in the last term arises from the discontinuities of $\ln^3(m^2)$
according to~(\ref{discgen}). Now remark that none of the nonlogarithmic contributions in 
the {\em original} perturbative expression~(\ref{qqQCDpert}) contribute to the spectral density. 
Thus similarly all subtraction
terms in~(\ref{s_i}), which are necessary for RG invariance of the original expression, 
do not contribute as well, thus making the final expression to be optimized relatively simpler. 

This point is worth elaborating in some detail. Just as for the GN model, instead of using the spectral density
we could  apply the RGOPT method more directly to the original perturbative expression of the condensate, 
Eq.~(\ref{qqQCDpert}), including in this case 
the subtractions~(\ref{sub}),~(\ref{s_i}) required by RG invariance (and also removing an overall factor $m$
from Eq.~(\ref{qqQCDpert}) to define a nontrivial $ \la \bar q q\ra$ in the chiral limit). When this is done, 
one finds rather unstable values for the optimized mass, coupling, and resulting condensate, 
showing no clear empirical convergence pattern at successive orders, at least
at the presently available (three-loop) order. 
Furthermore these results tend to give a wrong sign (positive, or ambiguous) condensate. 
More precisely, at the first nontrivial
$\delta^0$ (one loop) order, firstly there is no common nontrivial RG and OPT Eqs. solution.  
Considering then the OPT or RG Eqs. alone, both give a  positive condensate, of roughly 
the right order of magnitude: 
$\la \bar q q\ra(\delta^0)= \sqrt{e}/(2\pi^2) \Lam_{\ms}^3 \simeq 0.08 \Lam_{\ms}^3$ from the OPT, 
and a very similar value is obtained from the RG. Next at the 
$\delta^1$ (two-loop) order, the (unique) AF-compatible branch solution of the combined RG and OPT Eqs. 
(\ref{RGred}), (\ref{OPT})
gives complex-valued optimized coupling, mass and condensate, with a negative real part for the condensate 
but a much larger imaginary part: $\la \bar q q\ra(\delta^1)\simeq (-0.08 \pm 0.37\, i)  \Lam_{\ms}^3$, 
a result clearly ambiguous. These calculations are also not very stable upon 
different truncations of perturbative higher order terms in the RG equation (\ref{RGred}).
Furthermore, attempting to recover real AF-compatible
solutions by a perturbative renormalization scheme change, both at orders $\delta$
and $\delta^2$, happens to give no solutions 
(unlike for the pion decay constant case where the imaginary parts were small enough to allow for such 
a scheme change with very stable results~\cite{rgopt_alphas}).\\ 
We can trace this wrong sign and unstable behavior 
to the fact that in four dimensions the (presumably dominant) one-loop contribution  to the
fermion condensate, given by the very first graph in Fig.~\ref{qqpert}, 
is quadratically divergent. The contribution of this quadratic divergence 
has actually the correct negative sign.
Incidentally, in the Nambu--Jona-Lasinio model~\cite{NJL}, an effective cut-off handles this
divergence, and the quark condensate has automatically the correct sign.
Note that in the NJL model the condensate (or equivalently the mass gap 
in the widely used leading order large-$N$ approximation) 
is precisely given by the very same
one-loop first graph of Fig.~\ref{qqpert}, up to trivial overall factors, while genuine QCD contributions only enter at the next 
orders with gluon and further quark loop dressing.
In dimensional regularization, at lowest orders in the coupling,
one finds that the extra subtractions~(\ref{sub}),~(\ref{s_i}) have a sign opposite 
to the sign of the similar terms in the GN model, and that this is due to
the fact that the pole of $\Gamma(1-D/2)$ in the perturbative calculation of the condensate 
changes sign when going from dimension $D=2$ (corresponding to the logarithmic divergence in the GN model,
and quadratic divergence in the $D=4$ NJL model) 
to dimension $D=4$ (corresponding to QCD). Indeed, as is well-known most of the phenomenological 
successes of the NJL model rely
strongly on the physical cutoff interpretation of the quadratically divergent mass gap in four 
dimensions\footnote{The NJL model may be formulated in dimensional regularization to some extent, 
in dimensions $2 < D < 4$, but with rather odd properties,  see {\it e.g.}~\cite{NJLdimreg}}. 

Hence, the RGOPT appears bound to lead to a wrong sign and/or unstable results, if 
applied directly to the QCD perturbative 
expression of the quark condensate evaluated in dimensional regularization and related $\ms$ scheme at low orders. 
We do not know whether higher orders would cure this problem by ultimately stabilizing the result,
but this appears rather unlikely since the first few orders are likely to remain dominant
in our approach. (Indeed a general property of the optimized perturbation 
is that the optimized coupling $\t g$ turns to be reasonably small, so that the first few orders
dominate). We did not have this problem in our previous works~\cite{rgopt1,rgopt_alphas,rgopt_Lam} 
in which we were dealing with the pole mass and the pion decay constant, which are 
only logarithmically divergent quantities.
Yet one should not hastily conclude that the OPT or  
RGOPT approaches are bound to fail in any situation where quadratic divergergences would be present
in a cutoff regularization\footnote{In fact the (standard) OPT has been applied in the
framework of the effective NJL model with a cut-off at next-to-leading $\delta$ order~\cite{NJLOPT}, 
giving sensible results beyond the large-$N$ approximation, and  
consistent with important basic properties like the Goldstone theorem, the GMOR relation, etc.}. 
Rather, the above problems stress that in a given model it is crucial to choose 
carefully the basic entity to be perturbatively modified and optimized within the RGOPT framework.
(This is analoguous to the traditional variationnal Rayleigh-Ritz method in quantum mechanics, 
where the trial wave functions should often be appropriately chosen to obtain a sensible 
result). 
This is why for QCD one must use instead the spectral density, which in our framework we anyway
derived from the very same original perturbative condensate expression~(\ref{qqQCDpert}), but which at 
the same time formally gets rid of the influence of quadratic 
divergences. 
Indeed, only the infrared part $\lambda \to 0$ in Eq.~(\ref{SD}) can 
generate a nonzero result in the chiral limit, which is thus insensitive to ultraviolet divergences~\cite{SDgen}.    
We note also that lattice evaluations of the condensate also bypass 
this potential quadratic divergence problem by using the spectral density~\cite{qqlattSD,SDlatt_recent}, or by extracting
the condensate by more indirect methods, {\it e.g.} relying on the GMOR relation.\\

We thus proceed with the actual RGOPT calculations for the spectral density at successive perturbative orders.
First remark that, since there is no logarithmic $L_\lambda$ contribution in the spectral density 
at one-loop order (the one-loop $\ln m$ contribution in~(\ref{disc3}) only
giving the constant $1/2$), there is no nontrivial $\lambda\ne 0$ optimized solution 
of~(\ref{OPTlam})
at one-loop order. Thus we should start applying our method at the next two-loop order.
\subsection{Two-loop ${\cal O}(\delta)$ results}
Let us perform step by step the RGOPT optimization by restricting first~(\ref{SDqq3l}) 
at the first nontrivial two-loop order. Concerning the $\delta$-expansion given by (\ref{substlam}), 
it is crucial~\cite{rgopt_Lam,rgopt_alphas} 
to take the right value of the exponent $a$, determined by the lowest order
anomalous mass dimension, which makes the $\delta$-modified series 
compatible with RG properties and matching asymptotic freedom (AF), as we recalled in some details in Sec. III.B above. 
In the case of the large-$N$ limit of the GN model,
one has simply $a=\gamma_0/(2b_0)= 1$. Actually, since 
$m \la \bar q q\ra$ is RG invariant to all orders rather than $\la \bar q q\ra$, 
it is easily derived that the correct value to be used for $\la \bar q q\ra$, and 
thus for the related spectral density from (\ref{SD}), is 
\be
a=\frac{4}{3}(\frac{\gamma_0}{2b_0}).
\ee
Then to first nontrivial order in $\delta$ the modified series reads,
\be
-\rho^{\delta^1}_{QCD}=  \frac{3}{2\pi^2} \lambda^3 
\left(\frac{19}{58} +\frac{g}{\pi^2} ( L_\lambda -\frac{5}{12} )\right)\, ,
\label{rhod2l}
\ee
and the OPT~(\ref{OPTlam}) and RG~(\ref{RGlam}) equations
have a unique solution, using also~(\ref{BC}),
given in the first line of Table~\ref{tabn2}.
For a simpler first illustration we actually used the RG Eq.~(\ref{RGlam}) at the very first order 
with the one-loop coefficient $b_0$, in order to get simple analytic solutions.
Therefore we obtain $\la\bar q q\ra^{1/3}(n_f=2)(\mu\simeq 2.2\bar\Lam_2) \simeq -0.96\, \bar \Lam_2$,
a fairly decent value given this lowest nontrivial order. At two-loop perturbative order expression~(\ref{rhod2l}) does not depend explicitly of the number of flavors $n_f$, but a $n_f$ dependence enters into 
the optimized results indirectly from the RG Eq.~(\ref{RGlam}) involving $b_0(n_f)$, also entering $\bar\Lam_{n_f}$. 
The corresponding optimized coupling
is $\t \alpha_S\equiv \t g/(4\pi)\simeq 0.83$, a moderately large value 
very similar to the optimal coupling values obtained, at first nontrivial RGOPT order, 
when considering the pion decay constant $F_\pi$ in ref.~\cite{rgopt_alphas}. 
Of course the precise number obtained for the condensate depends on the precise 
definition of the $\bar\Lam$ reference scale, which is generally perturbative 
and a matter of convention. To get the numbers in the first lines of Table \ref{tabn2} we have used
the simpler one-loop form, $\bar\Lambda=\mu e^{-1/(2b_0 g)}$, consistently with the one-loop RG Eq. used. 
When comparing below more precisely with other phenomenological
determinations of the condensate, we will use a more precise perturbative 
definition of $\bar\Lam$, at four-loop order, in agreement with
most other present determinations. Remark also that the condensate being scale dependent,
our RGOPT optimization fixes also a scale, consistently with a defining
convention for $\bar \Lam$, as indicated in Table \ref{tabn2}. \\
For $n_f=3$ at order $\delta$ one finds similarly the optimized results given in
the first line of Table
\ref{tabn3}, 
indeed very close to the $n_f=2$ results above. 

Now, since our basic expression originated from an exact two-loop
calculation, it is {\it a priori } more sensible to apply the RG Eq.~(\ref{RGlam}) 
at the same two-loop order, in order to capture
as much as possible higher-order effects. Doing this we obtain the results given in the second lines of 
Tables~\ref{tabn2},~\ref{tabn3} for $n_f=2, 3$. 
Those results are therefore to be considered more consistent at two-loop order. One can already
observe the substantial decrease of the optimal coupling $\alpha_S$ to a more perturbative value,
and the correspondingly higher optimal scale $\mu$, with respect to the results using pure one-loop RG equation. 

For completeness and later use
we also give in Table~\ref{tabn2},~\ref{tabn3} the corresponding values of the {\em RG invariant} condensate
$\la\bar q q\ra_{RGI}$, perturbatively defined in our normalization as
\be
\la\bar q q\ra_{RGI} = \la\bar q q\ra(\mu)\,(2b_0\,g)^{\frac{\gamma_0}{2b_0}}\left(1+
(\frac{\gamma_1}{2b_0} -\frac{\gamma_0\,b_1}{2b^2_0})\,g +{\cal O}(g^2)\right),
\label{qqRGI}
\ee
where higher order terms not shown here are easily derived from integrating 
$\exp{[\int\, dg \gamma_m(g)/\beta(g)]}$
thus know perturbatively to four-loop $g^3$ order 
since only depending on the RG function coefficients~\cite{bgam4loop} $b_i, \gamma_i$. 
The factor multiplying the scale-dependent condensate $\la\bar q q\ra(\mu)$ in (\ref{qqRGI}) 
is obviously the inverse of the one defining similarly a scale-invariant mass, given explicitly 
to four-loop order in the literature (see {\it e.g.}~\cite{QCDruncode}). We will calculate 
the RG invariant condensate at successive orders in Tables~\ref{tabn2},~\ref{tabn3} using~(\ref{qqRGI})  
consistently at the same perturbative order as the RG order used in Eq.~(\ref{RGlam}), 
and taking $g\equiv \t g=4\pi \t \alpha_S$, the corresponding
optimized values obtained at each order. (Alternatively we could optimize directly the expression~(\ref{qqRGI}) instead of $\la\bar q q\ra(\mu)$ (with 
the last term $\propto \gamma_m(g)$ in the RG Eq.~(\ref{RGlam}) removed): This
gives the same optimized solutions, as expected since it is formally 
completely equivalent, up to tiny numerical differences due to perturbative reexpansions,  
of less than $10^{-3}$ relative to the numbers given in Tables~~\ref{tabn2}, \ref{tabn3}.)
\begin{table}[h!]
\begin{center}
\caption[long]{Main optimized results at successive orders for $n_f=2$, for the optimized spectral 
parameter $\t \lambda$, the  
optimized coupling $\t \alpha_S$, and resulting optimized condensate. We also give the RG invariant condensate
 $\la\bar q q\ra^{1/3}_{RGI}$ calculated at the consistent perturbative order from~(\ref{qqRGI}). $\bar\Lam_2$
 is conventionally normalized everywhere by Eq.~(\ref{Lam4pert}), except in the very first line
 where the one-loop expression $\bar\Lam\equiv \mu\,e^{-1/(2b_0\,g)}$ is rather used.} 
\begin{tabular}{|l||c|c|c|c|c|}
\hline
$\delta^k$, RG order  &  $\ln \frac{\t \lambda}{\mu} $ & $\t \alpha_S$ & 
$ \frac{-\la\bar q q\ra^{1/3}}{\bar\Lam_2}(\t \mu)$ &
$ \frac{\t \mu}{\bar\Lam_2}$ & $\frac{-\la\bar q q\ra^{1/3}_{RGI}}{\bar\Lam_2}$ \\
\hline
$\delta$, RG 1-loop & $-\frac{2275}{10092}$ & $\frac{87\pi}{328}\simeq 0.83$ & $ 0.962$ & $ 2.2$ & $0.996$ \\
$\delta$, RG 2-loop &   $-0.45 $    & $0.480$ & $ 0.822 $ & $ 2.8 $ & $0.821$ \\
\hline
$\delta^2$,  RG 2-loop  & $ -0.686 $ & $ 0.483 $ &$0.792$  & $2.797 $ & $0.792$ \\
$\delta^2$,  RG 3-loop  & $ -0.703 $ & $ 0.430 $  & $0.794$ &  $3.104 $ & $0.783$  \\
\hline
$\delta^3$, RG 3-loop  & $-0.838 $ & $0.405$ & $ 0.793$ & $3.306$ & $0.774$ \\
$\delta^3$, RG 4-loop  & $-0.820$ & $0.391$ & $0.796$   & $ 3.446$ & $0.773$ \\
\hline
\end{tabular}
\label{tabn2}
\end{center}
\end{table}
\begin{table}[h!]
\begin{center}
\caption{Same as Table \ref{tabn2} for $n_f=3$} 
\begin{tabular}{|l||c|c|c|c|c|}
\hline
$\delta^k$ order  &  $\ln \frac{\t \lambda}{\mu} $ & $\t \alpha_S$ & 
$ \frac{-\la\bar q q\ra^{1/3}}{\bar\Lam_3}(\t \mu)$ &
$ \frac{\t \mu}{\bar\Lam_3}$ & $\frac{-\la\bar q q\ra^{1/3}_{RGI}}{\bar\Lam_3}$ \\
\hline
$\delta$, RG 1-loop &   $-\frac{283}{972}$    & $\frac{27\pi}{104}\simeq 0.82$ & $0.965 $& $2.35$ & $0.987$ \\
$\delta$, RG 2-loop &   $-0.56 $    & $0.474$ & $0.799 $ & $ 3.06$ & $0.789$ \\
\hline
$\delta^2$,  RG 2-loop  & $ -0.766 $ & $ 0.493 $ &$0.776$  & $2.942$ & $0.772$ \\
$\delta^2$,  RG 3-loop  & $ -0.788 $ & $ 0.444 $  & $0.780$ &  $3.273$ & $0.766$ \\
\hline
$\delta^3$, RG 3-loop  & $-0.967 $ & $0.414$ & $ 0.769$ & $3.540$ & $0.745$\\
$\delta^3$, RG 4-loop  & $-0.958$ & $0.400$ & $0.773$   & $ 3.700$ & $0.744$ \\
\hline
\end{tabular}
\label{tabn3}
\end{center}
\end{table}
%
\subsection{Three-loop ${\cal O}(\delta^2)$ results}
At three-loop $g^2$ order the $n_f$ dependence enters explicitly within the perturbative expression of 
the spectral density, see Fig~\ref{qqpert} and the last 
$g^2$ coefficient in Eq.~(\ref{SDqq3l}),~(\ref{q3lSD}). 
This is interesting also in view
of other results on the variation of the condensate value with the 
number of flavors~\cite{qqflav,LattFLAG}.\\ 

We find a unique real AF-compatible optimized solution. More precisely, 
at this three-loop order there are actually 
two real optimized solutions for $\t L_\lambda, \t\alpha_S$, but the selection of the right physical solution
is unambiguous since only one is clearly compatible
with AF behavior for $g=4\pi\alpha_S\to 0$, $\ln (\t \lambda/\mu) \simeq -d_k/(2b_0 g) +{\cal O}(1)$ with 
$d_k={\cal O}(1)$, 
both for the RG and OPT equations.\footnote{Due to the nontrivial relations  
between $\ln m$ and $\ln \lambda$, Eqs.~(\ref{discgen}), (\ref{disc3}), 
the  $1/g$ coefficient of the correct AF branch
$L_\lambda(g)$ for $g\to 0$
is not exactly $-1/(2b_0)$, like it is for $\ln m/\mu$ (Eq.~(\ref{rgasympt})), essentially determined by the leading logarithms
$g^k \ln^k (m/\mu)$ behavior.
But this $1/g$ coefficient has the correct AF sign, being at order $\delta^k$: $-d_k/(2b_0)$ with 
$d_k>0$ a constant close to $1$, which approaches slowly $d_k\to 1$ as the perturbative order $k$ increases.} 
In contrast the other real solution has for $g \to 0$ a coefficient 
of opposite sign to AF, and gives $\t L_\Lambda =\ln \t \lambda/\mu >0$, which also means 
incompatibility with perturbativity, since we expect $\mu \gg \t \lambda$  just like the perturbative
range being  $\mu \gg \t m \sim \bar\Lam$ for the original expansion with mass dependence.
Explicitly we obtain for $n_f=2, 3$ the results given in the third and fourth lines of 
Tables~\ref{tabn2},~\ref{tabn3} respectively. More precisely as indicated there 
the third line results were obtained by taking the RG Eq.~(\ref{RGlam}) truncated at two-loop order,
while the fourth lines were obtained by taking the full three-loop RG equation. These results exhibit a very good
stability when confronted with the relative arbitrariness in the order of the RG equation.
Of course it is more legitimate to use the three-loop RG equation consistently at this three-loop order,
that we shall adopt for our final determination of the condensate.
Notice also that the optimal
values $\t \alpha_S$ decreased by almost a factor two with respect to the lowest nontrivial order result above,
which indicates that the resulting series is much more perturbative. But $\t \alpha_S$ almost
does not change as compared to the more consistent two-loop results. 
Similarly $\t \alpha_S$ further slightly
decreases and the optimal scale $\t \mu$ increases when going from two to three loops in the RG Eq.~(\ref{RGlam}).\\ 

As a matter of numerical detail, to obtain the results in Tables~\ref{tabn2},~\ref{tabn3} we took a convention of
the QCD scale $\bar\Lam$ based on a perturbative four-loop expression~\cite{PDG}:
\be
 \ds \bar \Lam_{n_f}^{4-loop}(g) \equiv 
\mu  \,e^{-\frac{1}{2b_0\,g}}\,(b_0\:g)^{-\frac{b_1}{2b^2_0}}  \exp \left[ -\frac{g}{2b_0} \cdot  
 \left((\frac{b_2}{b_0}-\frac{b^2_1}{b^2_0})   
 +(\frac{b^3_1}{2b^3_0} -\frac{b_1 b_2}{b_0^2} +\frac{b_3}{2b_0}) g\right) \right]\;.
\label{Lam4pert}
\ee
This is convenient and important to compare precisely below 
with most recent other determinations,
using the same four-loop perturbative order conventions for $\bar\Lam$. 
Actually when performing the RG Eq.~(\ref{RGlam}) at order $\delta^k$ it would be more natural to  
adopt a $\bar\Lam$ convention at the consistent $(k+1)$-loop order
$\bar\Lam_{k+1}$, given from (\ref{Lam4pert}) by taking $b_3=0$ (and $b_2=0$) respectively at three loops (two loops).
But this only affects an overall normalization of the final result, as $\bar\Lam$ itself is not involved 
in the actual optimization process when performing Eq.~(\ref{OPT}) and~(\ref{RGlam}). Besides, starting 
at three-loop order the differences obtained from such different conventions are minor. 
(The $\bar\Lam$ convention also affects the precise value of the optimal scales $\t \mu/\bar\Lam$ 
in Tables~\ref{tabn2},~\ref{tabn3} from which we shall start evolution 
to higher scale to compare with other determinations
in the literature, see below). Strictly speaking, the different values of $\la\bar q q\ra(\mu)$
obtained in Tables~\ref{tabn2},~\ref{tabn3} {\it e.g.} at three-loop order cannot be directly compared, being obtained at different scales $\t \mu$. Thus we also give the scale-invariant condensate values 
$\la\bar q q\ra_{RGI}$ which can be more appropriately compared. 

Notice also that in spite of the more than 10\%
change in the optimal coupling $\t \alpha_S$ when taking two- or three-loop RG, 
the final physical value of the condensate only varied
by $0.25\%$\footnote{This result appears so stable partly due to the $1/3$ power 
of the condensate in Tab.~\ref{tabn2},~\ref{tabn3}. For the actual optimization performed on 
the quantity $\la\bar q q\ra$, the corresponding variation is $\simeq 0.7\%$.}: 
This also reflects a strong stability. 
Moreover the value of $\la\bar q q\ra^{1/3}/\bar\Lam$ changed by about $20\%$ with respect to the
crude two-loop order result (first lines of Tables), but much less when compared to the more consistent
two-loop order result (second lines). This 
shows {\it a posteriori} that stability appears at the first nontrivial two-loop result, with an 
already quite realistic value. This stability, suggesting that we remain within the domain of validity of perturbation theory, 
is an important requirement for the usefulness of our method.
A similar behavior was observed
when optimizing the pion decay constant in~\cite{rgopt_alphas}.
Remark also that the optimized coupling values $\t \alpha_S$ at successive orders happen to be rather 
close to those for which the scale invariance factor in~(\ref{qqRGI}) multiplying 
$\la\bar q q\ra(\mu)$ would be exactly one 
(which for $n_f=2$ happens for $\alpha_S\simeq 0.483, 0.461, 0.457$ respectively at two-, three- and  
four-loop order). In other words, $\t \alpha_S$ is close to a (variational) 
``fixed-point'' scale-invariant behavior. Had we found 
optimized $\t \alpha_S$ values a factor 2-3 smaller, or larger, we would obtain no valuable results beyond 
ordinary perturbation in the first case, or much more unstable results in the second case.  
However we stress that the optimal coupling 
$\t \alpha_S$ or optimal mass $\t m$  do not have really a universal physical interpretation since the precise
$\t \alpha_S$ and $\t m$ values depend on the physical quantity being optimized. For instance when optimizing $F_\pi$ in 
ref.~\cite{rgopt_alphas} at a given perturbative order, the corresponding $\t \alpha_S$ 
values were pretty much close to the present ones, but nevertheless slightly different.
The physically meaningful result is obtained when replacing
$\t \alpha_S$ and $\t \lambda$ within the quantity being optimized, like here the condensate. \\

Comparing Tables~\ref{tabn2},~\ref{tabn3} it also clearly appears that the {\em ratio} of 
the quark condensate to $\bar\Lam^3$ has a moderate dependence 
on the number of flavors $n_f$, although there is a definite trend
that $\la\bar q q\ra^{1/3}_{n_f=3}$ is smaller by about $2-3\%$ with respect to 
$\la\bar q q\ra^{1/3}_{n_f=2}$, in units of $\bar\Lambda_{n_f}$,  
at the same perturbative orders. The smallness of this difference
was quite expected, due to the $n_f$ dependence only appearing
at three-loop order and the overall stability of the modified perturbation.
However from various different estimations, including lattice~\cite{Lamlatt}, 
and ours~\cite{rgopt_alphas}),
there are some indications that $\bar \Lam_2 > \bar\Lam_3$ 
(although unclear from uncertainties, due to a larger uncertainty on $\bar \Lam_2$), which therefore
could indirectly further affect the actual flavor dependence of the condensate. 
We shall come back in more detail below on this point in the phenomenological discussion in next section, 
after establishing our final result for the precise condensate values. 
\subsection{Four-loop ${\cal O}(\delta^3)$ results}
We finally consider the optimization of the spectral density at four-loop order, the maximal
order available at present. In fact, the complete standard perturbative expression of our starting 
expression for the condensate, 
{\it i.e.} the next $\alpha_S^3$ order correction to~(\ref{qqQCDpert}), is not fully known at present. But
it obviously takes the form
\be
m\,\la \bar q q \ra_{QCD}^{4-loop}(m,g) =
\frac{3}{2\pi^2}\,m^4 \: (\frac{g}{16\pi^2})^3\;
\left( c_{40} L^4_m + c_{41} L^3_m +c_{42} L^2_m +c_{43} L_m + c_{44} \right) 
\label{qq4l}
\ee
where $L_m\equiv \ln (m/\mu)$ and we choose a convenient overall normalization with respect to lowest order terms in~(\ref{qqQCDpert}).
Now the leading (LL), next-to-leading (NLL) and next-to-next-to-leading (NNLL) logarithms coefficients
$c_{40}-c_{42}$ are easy to derive from RG invariance properties as being fully determined by lowest orders.
The NNNLL $\ln m$ coefficient $c_{43}$ can also be inferred by RG  properties from the available 
anomalous dimension of the vacuum energy, calculated by Chetyrkin and Maier~\cite{vac_anom4}, and related to $s_3$
given in Eq.~(\ref{s_i}). Explicitly, we obtain
\bea
&& c_{40} \simeq  4836.74 \:(4533.33) \nn \\
&& c_{41} \simeq -12282.5 \:(-11292.4) \nn \\
&& c_{42} \simeq  15606.4 \:(12648.1) \nn \\
&& c_{43} \simeq -18588.6 \:(-15993.5) 
\label{c4i}
\eea
where the first and second numbers correspond to $n_f=2$ and $n_f=3$ respectively. (N.B.: We can obtain
the generic algebraic values of $c_{4i}(n_f)$ but these are rather involved and not particularly instructive,
so we prefer to keep an approximate numerical form for the relevant $n_f=2, 3$ case in~(\ref{c4i}).)
Thus only the nonlogarithmic coefficient $c_{44}$ is actually unknown at present,
and could be quite challenging to compute. But since the nonlogarithmic parts cannot 
contribute to the spectral density,  the latter can thus be fully determined
at four loops! This gives 
for the {\em exact} perturbative four-loop contribution to the spectral density, after taking the logarithmic
singularities according to Eqs.~(\ref{disc}),(\ref{discgen}):
\be
-\rho^{\ms}_{\mbox{4-loop}}(\lambda) =  \frac{3}{2\pi^2}\, \lambda^3 \: (\frac{g}{16\pi^2})^3\;
\left( c_{40}(n_f) (2 L^3_\lambda-\frac{\pi^2}{2} L_\lambda) +c_{41}(n_f) 
(\frac{3}{2}  L^2_\lambda -\frac{\pi^2}{8}) + c_{42}(n_f) L_\lambda +\frac{1}{2} c_{43}(n_f) \right) \, ,
\label{q4lSD}
\ee
to be added to the three-loop expression in~(\ref{SDqq3l}). 
It allows us to calculate the spectral density and the related condensate  
at three successive orders of the variationally modified perturbation, 
which gives further confidence and an important
stability and convergence check of our result. 

We obtain at four-loop order once more a unique real common RG and OPT AF compatible solution.
(The brute optimization results actually give several real solutions for $\t \lambda, \t \alpha_S$ but
there are no possible ambiguities since all solutions are eliminated from the AF compatibility 
requirement, except a single one, with $\t \alpha_S >0$ and $\t L_\lambda <0$ as expected.) 
Explicitly we obtain  the optimization results given in the 5th and 6th lines of Tables~\ref{tabn2},~\ref{tabn3}
respectively for $n_f=2, 3$, 
where to illustrate the stability 
the 5th lines correspond to taking the RG Eq.~(\ref{RGlam}) at three-loop order, and the 6th lines 
(more consistently) at four-loop order ($\bar\Lam$ being always taken now 
at four-loop order from Eq.~(\ref{Lam4pert})). 
One observes a further decrease of the optimal coupling $\t \alpha_S$ to more
perturbative values, with respect to the three-loop results above, as well as the corresponding decrease
of $\t L_\lambda$, meaning that $\t \mu$ is also larger.  The stabilization/convergence of the
results is even clearer for the scale-invariant condensate $\la\bar q q \ra_{RGI}$ given in the last columns
in Tables~\ref{tabn2},~\ref{tabn3}, which at four-loop order has almost no variation upon RG Eq. truncations.

To better appreciate the very good stability of these results, consider the basic perturbative 
expression of the condensate~(\ref{SDqq3l}) up to four loops 
in more numerical form (for $n_f=2$) and a more standard normalization of the coupling:
\bea
-\rho^{\ms}_{QCD}(4-loop)(\lambda) \simeq  && \frac{3}{2\pi^2}\,\lambda^3\:  
\left(-\frac{1}{2} + \frac{4\alpha_S}{\pi} (L_\lambda-0.42) + 
 (\frac{\alpha_S}{\pi})^2 (9.46 + (29.1 - 25.7 L_\lambda) L_\lambda) \right. \nn \\ 
 && \left. +(\frac{\alpha_S}{\pi})^3 (91.5 + L_\lambda (-129 + L_\lambda (-288 + 151 L_\lambda))\right)\;.
\eea
From this one can easily appreciate that the successive perturbative terms are not small, just like
in most perturbative QCD series: At successive orders the coefficients grow rapidly (even 
if partly damped by the decreasing $\alpha_S/\pi$ higher powers, provided that 
$\alpha_S$ remains rather moderate). 
In fact for the relevant above values of $\t \alpha_S\simeq 0.4-0.5$ roughly, 
and typically $\t L_\lambda\simeq -(.7-.8)$, 
depending on the RGOPT order, all successive perturbative
terms are roughly of the same order of magnitude. Now for the variationally modified perturbation
the successive sequences are quite different, but before any optimization the resulting series in $\alpha_S$
has perturbative coefficients that similarly grow at successive orders. But the RGOPT mass and coupling optimization
manage to stabilize the series, in such a way that the discrepancies between the three- 
and four-loop orders in the final $\la \bar q q\ra^{1/3}$ results are about $2\%$ or less. It is important
that the optimized sequence has thus clearly further stabilized from three- to four-loop order, 
to be more confident in a quite precise determination 
of the condensate, although the variation from the lowest nontrivial two-loop to three-loop results were 
already very reasonable, by only $\sim 4\%$. 

 It appears that these QCD RGOPT results
are more stable than the corresponding ones for the spectral density of the 
yet simpler large-$N$ GN model in Table~\ref{GNrho20}. This 
is a bit surprising {\it a priori }, given that direct optimizations, not going through the spectral density,  
give maximal convergence 
for the large-$N$ GN model~\cite{rgopt1}. In fact one can understand these results as follows. As explained
in section IV above, the rather slow convergence for the GN spectral density is entirely due to the 
large and growing factors of $\pi^{2p}$ coming from the discontinuities~(\ref{discgen}) at successive orders, 
spoiling the simple form of the series and `screening' the otherwise maximal convergence with 
the neat solution~(\ref{gLGN}). Now although the $\pi^{2p}$ coefficients from~(\ref{discgen}) are universal, 
thus the same for QCD, once combined with the original perturbative coefficients of~(\ref{qqQCDpert})
their {\em relative} contributions with respect to the other perturbative 
terms remain more reasonably of the same order in the QCD case than in the GN case. This is because 
the original perturbative coefficients are comparatively larger in the QCD case.
More precisely inspecting the QCD spectral density series 
at three-loop order $g^2$ in~(\ref{q3lSD}), the $\pi^2$ contribution (last term
in the RHS of~(\ref{q3lSD})) is roughly twice the other non-logarithmic contribution (first term
in the RHS of~(\ref{q3lSD})). Similarly at the next four-loop $g^3$ order from~(\ref{qq4l}) the $\pi^2$ contributions
are roughly twice the other non-logarithmic contribution ($c_{43}/2$). As long as those
$\pi^{2p}$ contributions remain roughly of the same order of magnitude as   
the original perturbative coefficients, such that some balance can occur from the optimization process, 
they should produce a moderate disturbance of the observed stability. We expect
these properties to remain true at even higher orders, because the QCD original perturbation coefficients
also grow rapidly with the order. \\

Lines 4 to 6 of Tables~\ref{tabn2},~\ref{tabn3}  give our direct optimization 
results for $n_f=2, 3$ and three or four loops respectively. 
To get a final result it could be legitimate to take only the presumably more precise 
maximal four-loop perturbative order available, as is commonly done in most perturbative analysis. 
However to allow for a more realistic estimate of the theoretical error of our results, we will more
conservatively consider the difference between the three- and four-loop results (but 
using consistent RG perturbative order in each case)
as defining the theoretical uncertainty.
\section{Evolution to higher energy and phenomenological comparison}
In order to get a more precise result 
it is necessary to take into account the (moderate but not completely negligible) 
running of the condensate values, since the optimal scales
obtained are somewhat different at three- and four-loop order, though reasonably perturbative, 
roughly of order $\t \mu\gsim 1$ GeV. It is anyway necessary to perform a further evolution of scale 
if only to make contact with the more standard scale $\mu\simeq 2$ GeV where 
other (sum rules, lattice, etc.) 
condensate determinations are often conventionally given. 
\subsection{RGOPT $\la\bar q q\ra (\mu=2 \rm GeV)$ results for $n_f=2$ and $n_f=3$}
The procedure to evolve perturbatively
the condensate from one to another scale is straightforward since from exact RG invariance
of $m \la\bar q q\ra$ it is simply given by the inverse of the well-known running
of the quark masses,
\be
\la\bar q q\ra (\mu') =\la\bar q q\ra (\mu) \exp[{\int_{g(\mu)}^{g(\mu')} dg 
\frac{\gamma_m(g)}{\beta(g)} }]\, .
\label{running}
\ee
 Alternatively we may take the values of the scale-invariant condensate~(\ref{qqRGI})
as obtained in the last columns of  Tab.~\ref{tabn2},~\ref{tabn3} and extract from those
the condensate at any chosen (perturbative) scale $\mu'$ by using again~(\ref{qqRGI}) now taking $g \equiv 
4\pi\alpha_S(\mu')$. This is of course fully equivalent to performing the running from $\mu$ to $\mu'$
with Eq.~(\ref{running}).    
Since all relevant scales
$\t \mu$ obtained above are in a fairly perturbative range $\gsim 1$ GeV, we take a (four-loop) perturbative 
evolution\footnote{For $n_f=2$, it is implicitly understood that this evolution is performed in a simplified
QCD world where the strange and heavier quarks are all infinitely massive, {\it i.e.} `integrated out'. Otherwise
it would not make sense perturbatively to take into account the strange quark mass threshold effects 
on the running. For $n_f=3$ we can perform a more realistic running, taking into account properly the charm
quark mass threshold effects on $\alpha_S(\mu\sim m_c)$, see below.}.
We choose the highest optimized scales obtained
(given by the four-loop results for both $n_f=2$ and $n_f=3$ cases) as reference low scale(s): 
$\mu_{ref}(2)=3.45\bar\Lam_2$ and $\mu_{ref}(3)=3.70\bar\Lam_3$ respectively. (N.B.: Given 
the present 
values of $\bar\Lam_3\simeq 340 \pm 8$ MeV\cite{PDG}, $\mu_{ref}(3)$ happens quite accidentally  
to be very closely below the charm quark mass threshold). 
For example this running gives a $\simeq 2\%$ increase
of the three-loop $\la\bar q q\ra^{1/3}_{n_f=2}$  value given in Tab.~\ref{tabn2} and quite similarly for $n_f=3$.
Putting all this together we obtain
\bea
&& -\la\bar q q\ra^{1/3}_{n_f=2}(\mu_{ref}(2)= 3.45\bar\Lam_2) =  (0.796-0.808) \bar\Lam_2, \nn \\
&& -\la\bar q q\ra^{1/3}_{n_f=3}(\mu_{ref}(3)= 3.70\bar\Lam_3) =  (0.773-0.796) \bar\Lam_3,
\label{qqn23fin}
\eea
which are our intermediate results in terms of $\bar\Lam$ and at the respective $n_f=2, 3$ 
optimal scales, including our estimated theoretical uncertainties, roughly of order $2\%$, given
by the range of differences between the three-loop results (evolved to the scales $\mu_{ref}$) and direct four-loop
results. We give results in the form~(\ref{qqn23fin}) in view of possibly more precise
determinations of $\bar\Lam_{2,3}$ in the future. 
Note that for both $n_f=2, 3$ the lowest values given 
in~(\ref{qqn23fin}) correspond to the presumably more accurate maximal four-loop results, 
which gave the $\mu_{ref}$ values directly from optimisation, so 
without possible extra uncertainties from running. \\

Next we 
perform a final evolution from the low reference scales $\mu_{ref}(2,3)$ relevant for $n_f=2$ and $n_f=3$
respectively as given in~(\ref{qqn23fin}), up to the conventional scale $\mu'=2$ GeV, where
from the present world average $\bar\Lam_3$ above we find $\bar\alpha_S(2\rm GeV)\simeq 0.305\pm 0.004$. 
For $n_f=3$ we take into account the four-loop expression of the perturbative matching \cite{matching4l} 
at the crossing of the charm quark threshold.
Overall this leads to an increase of the values in~(\ref{qqn23fin}) 
for $|\la\bar q q\ra|^{1/3}$ of about $\sim 4.6\%$ for $n_f=2$ and 
5.3\% for $n_f=3$ (in which taking into accout the charm quark threshold with matching
relations for $\alpha_S(\mu\simeq m_c)$ contributes
to $\sim -0.3\%$ with respect to a more naive one-step evolution ignoring charm threshold effects).
More precisely:
\bea
&& -\la\bar q q\ra^{1/3}_{n_f=2}(2 {\rm GeV}) =  (0.833-0.845) \bar\Lam_2 \nn \\
&& -\la\bar q q\ra^{1/3}_{n_f=3}(2 {\rm GeV}) =  (0.814-0.838) \bar\Lam_3\;.
\label{qqn232GeV}
\eea
To give a more precise determination for $n_f=2$ one obstacle is 
the presently not very precisely known value of the basic scale $\bar\Lam_2$. 
In principle it is beyond purely perturbative reach, as it cannot be `perturbatively connected'
to the more precisely known $\bar\Lam_3$ value~\cite{PDG}. Our own estimate~\cite{rgopt_alphas} 
of $\bar\Lam_2$ from the pion decay constant gave $\bar\Lam_2 \simeq 360^{+42}_{-30}$ MeV, while
some recent lattice determinations are $\bar\Lam_2\simeq 330\pm 45$ (staggered Wilson
fermions~\cite{Lamlatt}) and $\bar\Lam_2\simeq 331\pm 21$ (quark static potential method~\cite{LamlattVstatic14}).
(Incidentally the latter most precise present lattice determination tended to increase the central value 
of $\bar\Lam_2$ by $\sim 15$
GeV with respect to former similar determinations~\cite{LamlattVstatic}). Since lattice uncertainties are
mostly statistical and systematic, while ours are theoretical errors, it is not obvious how to combine all these
in a sensible manner. We thus prefer to keep separate estimates. For a representative illustration, 
combining our present results in~(\ref{qqn232GeV}) with the above quoted most precise lattice values
of $\bar\Lam_2$, we obtain
\be
-\la\bar q q\ra^{1/3}_{n_f=2}(2 {\rm GeV},\rm lattice\, \bar\Lam_2)\simeq  278 \pm 2 \pm 18 \, {\rm MeV};\;\;
\label{qqfin2latt}
\ee
where the first quoted error is our intrinsical theoretical error propagated from the one
in~(\ref{qqn23fin}), while the second larger uncertainty originates from the lattice ones on $\bar\Lam_2$.
Using instead solely our above quoted RGOPT determination~\cite{rgopt_alphas} of $\bar\Lam_2$ gives somewhat higher values
with larger uncertainties:
\be
-\la\bar q q\ra^{1/3}_{n_f=2}(2 {\rm GeV}, \rm rgopt\, \bar\Lam_2)\simeq 301 \pm 2 ^{+35}_{-25} \, {\rm MeV}.
\label{qqfin2rg}
\ee
For $n_f=3$ the more precisely known $\bar\Lam_3$ value from many different determinations allows
a more precise determination of the condensate. Taking the latest world average values~\cite{PDG}
$\bar \alpha_S(m_Z) =0.1185 \pm 0.0006$ translating in $\bar\Lam^{wa}_3=340\pm 8$ MeV, we obtain
\be
-\la\bar q q\ra^{1/3}_{n_f=3}(2 {\rm GeV},\rm \bar\Lam^{wa}_3)\simeq  281 \pm 4 \pm 7 \,{\rm MeV}\, 
\label{qqfin3wa}
\ee
where again the first error is our estimated theoretical uncertainty and the second one is from the world averaged
$\bar\Lam_3$.
Using alternatively solely our RGOPT determination~\cite{rgopt_alphas} of $\bar\Lam_3= 317^{+27}_{-20}$ MeV, 
gives instead slightly lower values but with larger uncertainties:
\be
-\la\bar q q\ra^{1/3}_{n_f=3}(2 {\rm GeV}, \rm rgopt\, \bar\Lam_3)
\simeq  262 \pm 4 ^{+22}_{-17} \, {\rm MeV}\,. 
\label{qqfin3rg}
\ee
Finally rather than fixing the scale from $\bar\Lambda$, it may be more sensible  
to give our results for the ratio of the scale-invariant condensate with another physical scale, which is
a parameter-free prediction. Taking the $\la\bar q q\ra^{1/3}_{RGI}$ results in Tab.~\ref{tabn2},~\ref{tabn3}, and using solely our previous RGOPT results~\cite{rgopt_alphas} for $F/\bar\Lam_2$ 
and $F_0/\bar\Lam_3$ (where $F$ ($F_0$)
are the pion decay constant for $n_f=2$, $n_f=3$ respectively in the chiral limit), we obtain:
\be
\frac{-\la\bar q q\ra^{1/3}_{RGI,n_f=2}}{F}= 3.25\pm 0.02 ^{+0.35}_{-0.24}\;,
\label{finRGIn2}
\ee
where the first error comes from the present calculation of the condensate while 
the second ones from taking the most conservative range combining linearly three- and four-loop order uncertainty 
results for  $F/\bar\Lam_2$ from Eq.~(4.28)
of \cite{rgopt_alphas}. A less conservative estimate may be obtained alternatively 
by taking the range spanned by the maximal
available four-loop results for $F/\bar\Lam_2$ correlated with the four-loop condensate results. This gives
\be
\frac{-\la\bar q q\ra^{1/3}_{RGI,n_f=2}}{F}= 3.26\pm 0.01 ^{+0.22}_{-0.16}\;.
\ee
Similarly for $n_f=3$ we obtain:
\be
\frac{-\la\bar q q\ra^{1/3}_{RGI,n_f=3}}{F_0}= 3.04\pm 0.04 ^{+0.14}_{-0.07}\;,
\label{finRGIn3}
\ee
where the first theoretical error comes from the condensate with the second one from the 
conservative range combining linearly three- and four-loop order uncertainty on 
$F_0/\bar\Lam_3$ from Eq.~(4.30) of \cite{rgopt_alphas}. 
As observed above, the direct results from optimization of $-\la\bar q q\ra(n_f)/\bar\Lam^3_{n_f}$ 
in Tables~\ref{tabn2},~\ref{tabn3} show a moderate relative decrease, of about $2-3\%$ only 
on $\la\bar q q\ra^{1/3}_{n_f=3}/\bar\Lam_3$. The effect appears slightly more pronounced, about $7\%$ relative reduction from $n_f=2$ to $n_f=3$
when comparing the central values of~(\ref{finRGIn2}) 
and~(\ref{finRGIn3}), due to the slight $4\%$ reduction of (central) $F_0$ relative to $F$, 
although this result is not clear, being affected with rather large
uncertainties. We may finally combine~(\ref{finRGIn2}) and~(\ref{finRGIn3}) to give
\be
\frac{\la\bar q q\ra^{1/3}_{RGI,n_f=3}}{\la\bar q q\ra^{1/3}_{RGI,n_f=2}} \simeq \left(0.97\pm 0.01\right)\,
\frac{\bar\Lam_3}{\bar\Lam_2} \simeq 
\left(0.94 \pm 0.01 \pm 0.12\right)\:\frac{F_0}{F} \simeq 0.86 \pm 0.01 \pm 0.11 \pm 0.05 \;,
\label{qq3to2}
\ee
where all our theoretical errors are combined linearly. In the results on the RHS of~(\ref{qq3to2})
the first quoted errors are the intrinsic RGOPT errors for the present condensate calculation only, and the second
larger one is propagated from the $F/\bar\Lam_2$ and $F_0/\bar\Lam_3$ RGOPT theoretical errors. 
We  also stress in~(\ref{qq3to2}) that our results are by construction in the strict chiral limit $m_q\to 0$. 
The result given for unspecified $F_0/F$ corresponds to the present sole RGOPT condensate estimate without 
extra input from other methods, while the last result uses the present lattice $F_0/F$ estimates~\cite{LattFLAG}
(with its own uncertainty $\sim 0.05$ quoted last). \\ 
\subsection{Comparison and discussion}
One may compare~(\ref{qqfin2latt}),~(\ref{qqfin2rg}) 
with the latest, presently most precise lattice determination, from the spectral 
density~\cite{SDlatt_recent} for $n_f=2$:
$\la\bar q q\ra^{1/3}_{n_f=2}(\mu=2 {\rm GeV}) =-(261 \pm 6 \pm 8) $, where the first error is statistical 
and the second is systematic. Our results in (\ref{qqfin2latt}) are thus compatible within uncertainties, 
though marginally if taking the other RGOPT determination of 
$\bar\Lam_2$ in~(\ref{qqfin2rg}) tending to be relatively high.
Note however that the above quoted lattice value from~\cite{SDlatt_recent} was 
obtained by fixing the scale with the
$F_K$ decay constant rather than using $\bar\Lam$ 
(moreover with $F_K$ being determined in the quenched approximation).
It is thus probably more judicious to compare our results 
for the RG invariant ratio~(\ref{finRGIn2}) with 
theirs~\cite{SDlatt_recent}: $2.77\pm 0.02\pm 0.04 $.
Overall, recent lattice determinations from various other methods lie
in the range roughly $\la\bar q q\ra^{1/3}_{n_f=2} \sim -(220-320)$ MeV for $n_f=2$~\cite{LattFLAG}, and quite similarly
for $n_f=3$. The most precise lattice $n_f=3$ determination we are aware of is 
$\la\bar q q\ra^{1/3}_{n_f=3}(2\rm GeV) =
-(245\pm 16)$ MeV~\cite{Lattqqn3}. Concerning the $n_f=3$ to $n_f=2$ condensate ratio, various 
lattice results have still rather large uncertainties at present~\cite{LattFLAG} 
but some recent results are compatible with a ratio unity~\cite{ss_uulattice}. 
Our results compare a bit better with the latest ones from spectral sum 
rules~\cite{qqSRlast}: $\la\bar u u\ra^{1/3} \sim -(276\pm 7)$ MeV (and for the ratio ~\cite{ss_uuSR,ss_uuSRlast}:
$\la\bar s s\ra/ \la\bar u u\ra = 0.74^{+0.34}_{-0.12}$). However 
the sum rules results~\cite{qqSRlast} actually determine precisely the current quark masses, extracting then 
the $\la\bar u u\ra$ value indirectly from using the exact GMOR relation~(\ref{GMOR}).\\

We stress again the rather moderate $n_f$ dependence of our result. This is in some tension with 
the larger estimated difference between $n_f=2$ and $n_f=3$ cases obtained by some authors~\cite{qqflav}. 
Since our results are by construction valid in the strict chiral limit, taken at face value 
they indicate that the possibly larger difference 
obtained by some other determinations is more likely due to the
explicit breaking from the large strange quark mass, rather than an intrinsic $n_f$ dependence property 
of the condensate in the exact chiral limit. 
\section{Summary and Conclusion}
We have adapted and applied our RGOPT method on the perturbative expression of the spectral density of the Dirac operator,
the latter being first obtained from the perturbative logarithmic discontinuities of the quark condensate in the $\ms$-scheme.
This construction allows successive sequences of optimized nontrivial results in the strict chiral limit at two-, three- 
and four-loop levels. These results exhibit
a remarkable stability and empirical convergence. The intrinsic theoretical error of the method, 
taken as the difference between three- and four-loop results, is of order $2\%$, 
while the final condensate value uncertainty is more affected
by the present uncertainties on the basic QCD scale $\bar\Lam$, specially with a larger uncertainty 
for $n_f=2$ flavors. The values obtained are rather compatible, within uncertainties, with the most recent lattice 
and sum rules determinations
for $n_f=2$, and our values indicate a moderate flavor dependence of the $\la\bar q q\ra^{1/3}_{n_f}/\bar\Lam_{n_f}$ ratio. 

\end{document}